\documentclass[]{bmcart}
\usepackage{amsthm,amsmath}
\RequirePackage[authoryear]{natbib}
\usepackage[utf8]{inputenc} 
\usepackage[T1]{fontenc}
\usepackage{graphicx}

\startlocaldefs
\endlocaldefs

\begin{document}

\begin{frontmatter}

\begin{fmbox}
\dochead{Full paper}


\title{Modelling of Geomagnetically Induced Currents in the Czech Transmission Grid}


\author[
  addressref={aff1,aff2},                   
  corref={aff1},                       
  email={svanda@sirrah.troja.mff.cuni.cz}   
]{\inits{M.}\fnm{Michal} \snm{\v{S}vanda}}
\author[
  addressref={aff3},
  email={anna.smickova@gmail.com}
]{\inits{A.}\fnm{Anna} \snm{Smi\v{c}kov\'a}}
\author[
  addressref={aff4},
  email={tatiana.vybostokova95@gmail.com}
]{\inits{T.}\fnm{Tatiana} \snm{V\'ybo\v{s}\v{t}okov\'a}}


\address[id=aff1]{
  \orgdiv{Astronomical Institute},             
  \orgname{Faculty of Mathematics and Physics, Charles University},          
  \street{Hole\v{s}ovi\v{c}k\'ach 2},
  \city{Prague 8},                              
  \postcode{CZ-18200},
  \cny{Czech Republic}                                    
}
\address[id=aff2]{%
  \orgname{Astronomical Institute of the Czech Academy of Sciences},
  \street{Fri\v{c}ova 298},
  \city{Ond\v{r}ejov},
  \postcode{CZ-25165},
  \cny{Czech Republic}
}
\address[id=aff3]{%
  \orgdiv{Department of Electroenergetics},
  \orgname{Faculty of Electrical Engineering, Czech Technical University},
  \street{Technick\'a 2},
  \postcode{CZ-16627},
  \city{Prague 6},
  \cny{Czech Republic}
}
\address[id=aff4]{
  \orgdiv{Department of Surface and Plasma Science},             
  \orgname{Faculty of Mathematics and Physics, Charles University},          
  \street{Hole\v{s}ovi\v{c}k\'ach 2},
  \city{Prague 8},                              
  \postcode{CZ-18200},
  \cny{Czech Republic}                                    
}



\end{fmbox}


\begin{abstractbox}

\begin{abstract} 
We investigate the maximum expected magnitudes of the geomagnetically induced currents (GICs) in the Czech transmission power network. We compute a model utilising the Lehtinen-Pirjola method, considering the plane-wave model of the geoelectric field, and using the transmission network parameters kindly provided by the operator. We find that the maximum amplitudes expected in the nodes of the Czech transmission grid during the Halloween storm-like event are about 15~A. For the ``extreme-storm'' conditions with a 1-V/km geoelectric field, the expected maxima do not exceed 40~A. We speculate that the recently proven statistical correlation between the increased geomagnetic activity and anomaly rate in the power grid may be due to the repeated exposure of the devices to the low-amplitude GICs.
\end{abstract}




\end{abstractbox}
%

\end{frontmatter}



\section{Introduction}
The Sun is our closest star. Not only it influences all massive bodies in the Solar system by its gravity, but it also affects the interplanetary space by electromagnetic forces. Solar magnetic field undergoes time variations on several times scales, the related phenomena are termed \emph{solar activity}. 

The solar activity affects the cosmic neighbourhood of our planet Earth. Charged particles arriving from the Sun interact with the Earth's magnetic field. These interactions yield the time variations of the geomagnetic field termed \emph{geomagnetic activity}. For instance, the severe activity spikes with rapid changes of the solar magnetic field occurring on time scales from seconds to minutes have to do with the explosive magnetic field reconnections in solar flares. The flares are usually connected with the coronal mass ejections \citep[CMEs;][]{Chen2011,Webb2012}, solar plasma hurled into the interplanetary space. When this plasmoid hits the Earth's magnetosphere, a geomagnetic storm occurs. 

The second type of disturbances that drive geomagnetic storms are the co-rotating interaction regions \citep[CIRs;][]{Heber1999}. These regions stem from a high-speed stream of the solar wind originating from a coronal hole and occur when the stable fast solar wind stream interacts with the surrounding slow solar wind in the low and middle latitude regions of the heliosphere. 

The space-weather effects are many. The perturbed ionosphere during the geomagnetic storm disturbs the propagation of radio waves that are transmitted through the ionosphere \citep[e.g.][]{Tsurutani2009}. This effect affects also the GPS accuracy \citep{Uray2017}. An overview of space weather-related issues was delivered, for instance, by \cite{Koskinen2001} or by \cite{Shea1998} in comprehensive summaries. 

The interactions between the solar-wind disturbances and the ionosphere cause the geomagnetic disturbances to propagate down to the ground. In turn, electric currents may be generated in conductive surfaces. The so-called geomagnetically induced currents (GICs) are induced in the ground, in seas, and also in various utility systems, such as electric power grids \citep{Pirjola2000}, oil and gas pipelines \citep{Pulkkinen2001}, and wired communication utilities, such as signalling networks \citep{Eroshenko2010}.

The effects are naturally expected to be larger in the countries close to the geomagnetic poles. For instance, \cite{Pulkkinen2003} studied the geomagnetic storm of 6--7 April 2000. During this event, large GICs were measured in technological systems, both in Finland and in Great Britain. By comparing the GIC measurements with the geomagnetic data from the magnetometer network, they identified and analysed the ionospheric drivers of large GICs during this event. Only during the sudden storm commencement at the beginning of the event, there were large GICs identified across Northern Europe with a coherent behaviour. Other peaks in GICs were registered during substorm intensifications, which were geographically localised. 

The Europe-wide modelling of GICs with a simplified grid \citep{Viljanen2014} demonstrated that in the years 1996--2008  the peaks were about 400~A in the Nordic countries, about 100~A in the British Isles, about 80~A in the Baltic countries, and less than 50~A in Central and South Europe. The maxima were seen during Halloween storms in 2003. However, \cite{Cid2014} stressed out that local magnetic disturbances seem to play a key role in assessing the potential risk factor of extreme events in specific regions. Therefore, the term “extreme storm” should not be associated with any threshold of any global geomagnetic index, and, consequently, the risks posed by GICs do not depend solely on the geomagnetic latitude. 

Also in countries considered ``GIC-safe'' until recently, the space weather adverse effects were studied. For instance, \cite{Bailey2017} modelled the expected GICs in the Austrian power grid. The model was later verified by \cite{Bailey2018} against the direct measurements. In Austria, in the region of the Alps, the ground resistivity is large, hence the GICs in the conductive utilities may be larger, with the expected effects to be equivalent to countries such as Denmark or Scotland. During the Halloween storms, the models expected a maximum of about 13~A in one of three studied nodes. 

A similar situation can be considered in Italy, where the Alps can be found in the north and the Apennines cover a significant portion of the country. \cite{Tozzi2019} showed that the GIC risk in Italy is moderate. Due to the low ground conductivity of the Alps, the north of Italy is more exposed to a possible damage to power grid components than the rest of Italy. The modelling with ``extreme-storm'' conditions of the electric field of 1~V/km demonstrated that one might expect GIC peaks of 97~A in the north. 

\citeauthor{Torta2012} modelled GICs in Spain. In the first study \citep{Torta2012}, they considered a segment of the national 400-kV transmission grid, which consisted of 17 nodes (substations) and 23 transmission lines. The model was validated against direct measurements during the geomagnetic storm that occurred on 24--25 October 2011. The comparison showed that the GIC model satisfactorily corresponds to the real measurements. The peak GIC values during this storm reached about 2~A at Vandellòs substation. The model predicted GICs of about 10~A for the 24 March 1991 storm for the Can Jardí station and, in general, the peak amplitudes up to 50~A for the extreme-storm conditions of the electric field of 1~V/km. In the follow-up study \citep{Torta2014}, they modelled the whole 400-kV Spanish power grid using the same method. They found that GIC amplitudes of tens of Amperes were not uncommon during the Halloween storms in 2003, with a maximum of 78~A at Mesa de la Copa substation. For the 1-V/km field, they performed the directional analysis and showed that in Spain, one could expect GICs of amplitudes of about 150~A.  

In the Czech Republic, the pioneering study on GIC was delivered by \cite{Hejda2005}. They modelled the geoelectric voltage on Dru\v{z}ba and IKL oil pipelines and compared these modelled voltages with the measured perturbations of the cathodic protection during Halloween storms. It was convincingly shown that the model and the measurements again corresponded to each other very well. 

Our study is motivated mainly by two recent projects. First, \cite{Vybostokova2019} investigated a statistical correlation between the occurrence of disturbances on the electric power transmission network in the Czech Republic and the geomagnetic activity described by the K index, which is typical for characterizing the level of geomagnetic activity in similar applications. The maintenance logs were provided by the power grid operators (both transmission and distribution lines). The maintenance logs were divided into 12 separate sets when the events were divided according to the affected device and voltage level. The events that could not be associated with GICs were excluded from those sets. We found that in the case of the data sets recording the disturbances on power lines at the high and very high voltage levels and disturbances on electrical substations, there was a statistically significant increase of anomaly rates in the periods of tens of days around maxima of geomagnetic activity compared to the adjacent minima of activity. There were hints that the disturbances were more pronounced shortly after the maxima than shortly before the maxima of activity. The results provided the first indirect evidence that the GICs may affect the occurrence rate of anomalies registered on power-grid equipment even in the mid-latitude country in the middle of Europe.

The study was followed by another study \citep{Svanda2020}, where we investigated the near-immediate effects of the exposure of the Czech power grid to space weather events. A superposed epoch analysis was used to identify possible increases in anomaly rates during and after geomagnetically disturbed days. It was shown that in the case of abundant series of anomalies on power lines, the anomaly rate increased significantly immediately (within 1 day) after the onset of geomagnetic storms. In the case of transformers, the increase of the anomaly rate was generally delayed by 2–-3 days. We also found that transformers and some electric substations seemed to be sensitive to prolonged exposure to substorms, with a delayed increase of anomalies. Overall, in the 5-day period following the commencement of geomagnetic activity, there was an approximately 5–10\% increase in the recorded anomalies in the Czech power grid and thus this fraction of anomalies is likely related to exposure to GICs.

Both studies provided us with hints or possible indirect evidence that GICs may play some role in the Czech electric power grid and affect the health of the key devices in the grid. Therefore, we were interested in what could be the amplitudes of the GICs expected in the Czech power grid. Our study uses simplified models to teach us about the possible GIC amplitudes in the power grid of the mid- to low-latitude country.

This study utilises the measurements of the geomagnetic field, which are used to obtain a model for the geoelectric field, and then the technological description of the Czech transmission network, which is essential for the proper modelling of the GICs. 

\subsection{Geomagnetic data}
We used the data from the nearest measuring station, the Geomagnetic Observatory Budkov in the \v{S}umava mountains, operated by the Institute of Geophysics of the Czech Academy of Sciences. They produce minute-by-minute measurements of the geomagnetic field vector. The measurements of the geomagnetic field were downloaded from the Intermagnet data archive. For the purpose of the study, we selected the period between 24 October  and 2 November 2003, which included the period of so-called Halloween storms. 

The geomagnetic storm began on 28 October with an arrival of a CME and lasted for more than 27 hours \citep[see an overview by][]{Gopalswamy2005}. It had many demonstrable effects on human infrastructures. For example, several satellites in space were highly affected \citep{Lopez2004}, the storm had effects even on the altitude control of spaceships \citep[e.g.][]{Huang2017}. The Solar and Heliospheric Observatory (SOHO) satellite failed temporarily. NASA's Advanced Composition Explorer (ACE) satellite experienced damage, and instruments aboard many spacecraft had to be shut down temporarily.

The aurora was visible from unusual places such as Florida or Texas in the US. The most significant effect of the Halloween storm of 2003 was the blackout that affected the Swedish power grid -- especially the city of Malm\"{o} \citep{Wik2009}. The GICs, that were generated and flowed in the transmission line, reached values of hundreds of ampers (the calculated maximum was 330~A just before the Malm\"{o} blackout). The reason why the blackout occurred was the excess heating in the power transformer and tripping of a 130-kV line by relays that had high sensitivity to the third harmonics that was the effect of the GICs. The blackout lasted for about one hour on the evening of 30 October and affected about 50\,000 customers in Sweden. There were also reported failures in South Africa transformers \citep{Gaunt2007}. 

The Halloween-storm period thus constitutes a very good testing interval to assess the susceptibility of the technological elements to GICs. It needs to be mentioned that this testing period was also selected in other studies, such as \cite{Hejda2005} or \cite{Bailey2017}. 

\begin{figure}[h]
  \includegraphics[width=\textwidth]{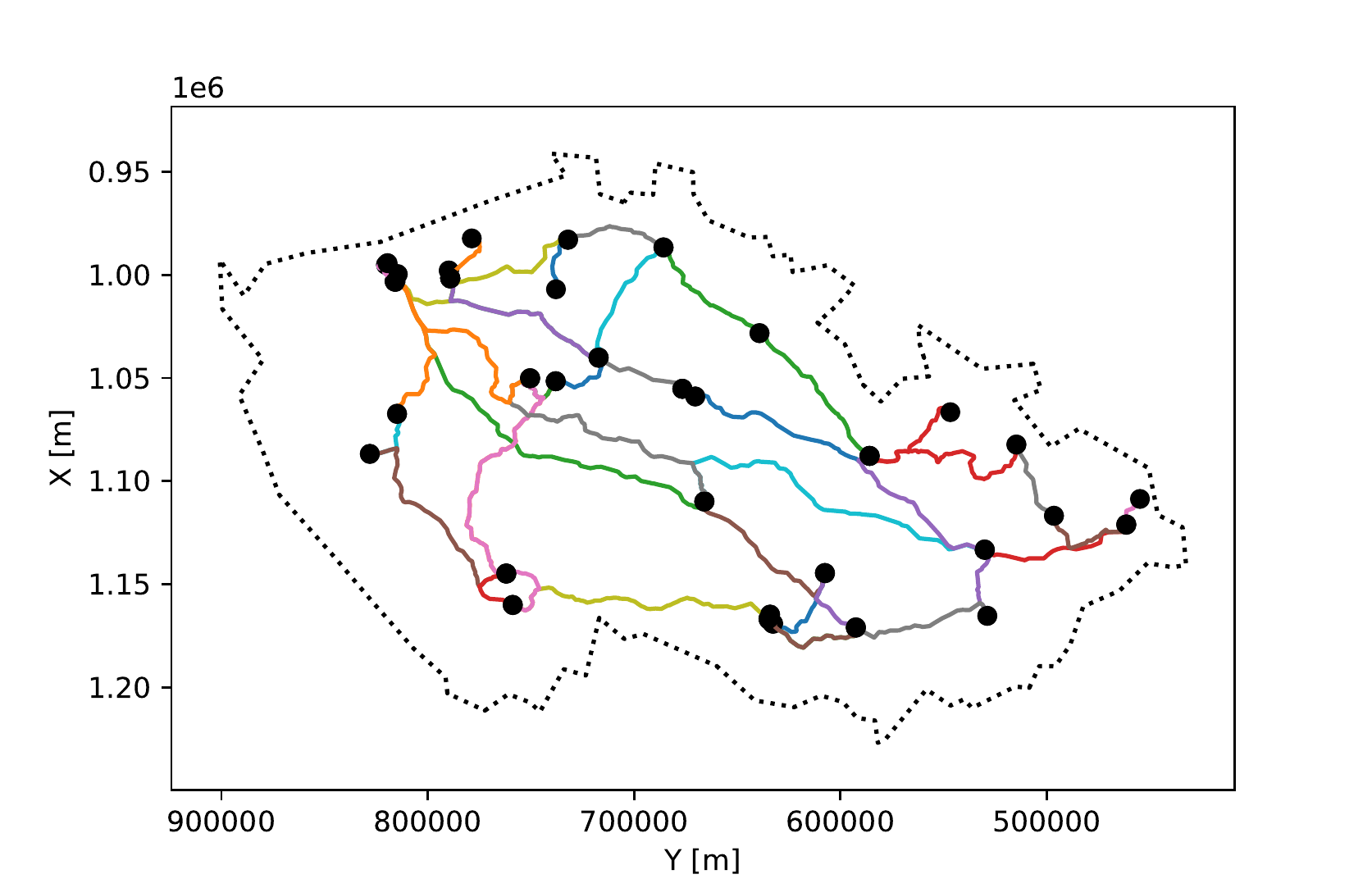}
  \caption{The map of the 400-kV Czech transmission network. The dotted line represents the simplified borders of the Czech Republic. Individual transmission lines are plotted with different colours. The units on both axes represent the S-JTSK coordinates. The black points indicate the location of the substations.}
  \label{fig:map}
\end{figure}
\section{Data}

\subsection{Transmission network data}

The Czech Republic is a central-European country, which is extended in the east--west direction (about~500 km length) compared to the ``width'' in the south--north direction (about  280~km). The Czech electric transmission grid is operated by the company \v{C}EPS, a. s. This company is owned by the state and maintains, restores, and develops 41 substations with 71 transformers around all the Czech Republic. The total length of lines constituting the spine of the transmission grid is 5502~km, from which the lines with a voltage level of 400~kV are 3867~km long, with a voltage level of 220~kV are 1824~km long, and the 110~kV transmission lines are 84~km long. The consumers are connected to the transmission network via the distribution networks, which are operated by three different private companies.

For our purpose, we focused on modelling of the GICs in the 400-kV transmission grid. We avoided using the cross-border lines because it was impossible to obtain the technical specifications of the substations in a territory of a different state. Our set incorporates 57 transmission lines and 37 substations. The longest line denoted as V420 is about 210~km long. The very high-voltage power-line labelling follows the logic of the letter V followed by a unique number, where the first digit indicates the voltage level (here the `4' indicates the 400-kV voltage level). 

For each line in question, we obtained the geographic coordinates of the tension towers and the resistivity of the line. The coordinates of the tension towers were given in the geographic S-JTSK coordinate system. This is a rectangular (Cartesian-like) coordinate system with the $X$ axis leading from the north to south and $Y$ coordinate from east to west. Thanks to the (small) size of the Czech Republic, the S-JTSK system is very close to the horizontal components of the Cartesian system in which the amplitudes of the geomagnetic field are given. The deviations of the Cartesian and S-JTSK systems in the Czech Republic are negligible for the purpose of this work. An overview map of the studied transmission network is given in Fig.~\ref{fig:map}.

For each of the substations, we obtained the geographic coordinates (again in the S-JTSK system) and the list of lines that connect there. For a few substations, we also obtained their grounding resistances, which were measured during the recent reconstructions (the values range from 0.031~$\Omega$ to 0.272~$\Omega$). For the remaining substations, the grounding resistances were not known, and thus we needed to estimate them by a generic value. For the favourable solution, we selected the average of the known resistances (which corresponds to 0.1~$\Omega$). 

\section{Methods}

\subsection{Model for geoelectric field}
In order to model the GICs in the power grid, we need to determine the geoelectric field first. Various approaches were applied to determine the geoelectric field from variations of the geomagnetic field, for example, a plane-wave method or complex image method \citep{Pirjola1998}. For the purpose of our study, we decided to use the plane-wave model. The disturbances caused by magnetospheric-ionospheric currents propagate vertically downwards as a plane wave, and disturbances caused by geoelectromagnetic variations are described as a wave in this approach. Our model showed good consistency with \cite{Hejda2005} where they modelled geomagnetically induced pipe-to-soil voltages in the Czech oil pipelines during the Halloween storms. 

A common assumption in similar studies of the geoelectric field is that the conductivity of the Earth only varies with depth (1D structure of the Earth). The Earth is replaced by a half-space with a flat surface, which is an acceptable approximation since GIC is a regional phenomenon. We are using a uniform ground resistivity model, which means that the Earth structure is regionally homogeneous with a constant conductivity $\sigma$. For the purpose of this study, we use the value of conductivity $\sigma=10^{-3}$ $\Omega^{-1}$\,m$^{-1}$ which is a typical value for Czech territory, and it was used also by \cite{Hejda2005}. Assumption of the sufficiently homogeneous geomagnetic field in the whole region of the Czech Republic let us to compute the geoelectric field only from its variations, and so we were not forced to consider ionospheric currents.

The plane-wave model yields the integrodifferential equation coupling electric and magnetic field, 
\begin{equation}
    E_y(t) = -\frac{1}{\sqrt{\pi \mu_0 \sigma}} \int_{0}^{\infty} \frac{g(t - u)}{\sqrt{u}} {\rm d}u
    \label{eq:E}
\end{equation}
in the time domain, where $g(t)= {\rm d} B_x(t)/{\rm d}t$. It is in agreement with causality, which means that at the time $t$, $E_y(t)$ depends only on the previous values of $g(t)$. The weight of affection by past values decreases with time. A stable solution can be achieved by integration over several hours. The geomagnetic coordinates are defined by the $x$ axis in the south to north direction and $y$ coordinates in the west to east direction. We note to the reader that compared to the S-JTSK coordinate system $(X,Y)$ of the tension towers and substations of the transmission grid, we have $x \parallel X$ and $y \parallel Y$, where each of them have the opposite senses. A similar equation to (\ref{eq:E}) can be written for $E_x$.

The square root in the denominator of (\ref{eq:E}) has a singularity at $t=0$. This problem is caused due to quasi-static approximation, where displacement currents are ignored. The solution to the problem exists, the details can be found in \cite{Love2014}. 

Since we approximated the Earth's surface as an infinite half-space, we are interested only in $x$ and $y$ components of the measured geomagnetic field. The time derivative of $\boldsymbol{B}=(B_x,B_y)$ can be discretized as

\begin{equation}
    \frac{\Delta \boldsymbol{B}(t_j)}{\tau}=\frac{\Delta \boldsymbol{B}(t_{j+1})-\boldsymbol{B}(t_j)}{\tau},
\end{equation}
where $\tau$ represents the data sampling ($\tau= 60$~s in our case). From the discrete values of $\boldsymbol{B}$ the induced electric field can be obtained by (numerical) convolution with the transfer function $\chi_R(t;\tau)$,

\begin{equation}
    \chi _R(t;\tau)=\frac{2}{\sqrt{\pi}}[\sqrt{t}H(t)-\sqrt{t-\tau}H(t-\tau)],
    \label{eq:transf}
\end{equation}
where $H(t)$ is the Heaviside function.

Then 

\begin{equation}
    \boldsymbol{C} \boldsymbol{E} (t_i;\sigma)=\frac{1}{\sqrt{\mu \sigma}} \sum_{j=1}^{i}\chi_R(t_i-t_j)\frac{\Delta \boldsymbol{B}}{\tau}(t_j),
    \label{eq:numgeoel}
\end{equation}
where 

\[
\textbf{C} = 
\begin{matrix}
    \begin{bmatrix}
        0 & -1\\
        1 &  0

    \end{bmatrix}
\end{matrix}
\]
is a spin matrix coming from the curl operator. Due to the plane wave approximation, $\boldsymbol{C}$ has only two dimensions.
\begin{figure}[h]
  \includegraphics[width=0.5\textwidth]{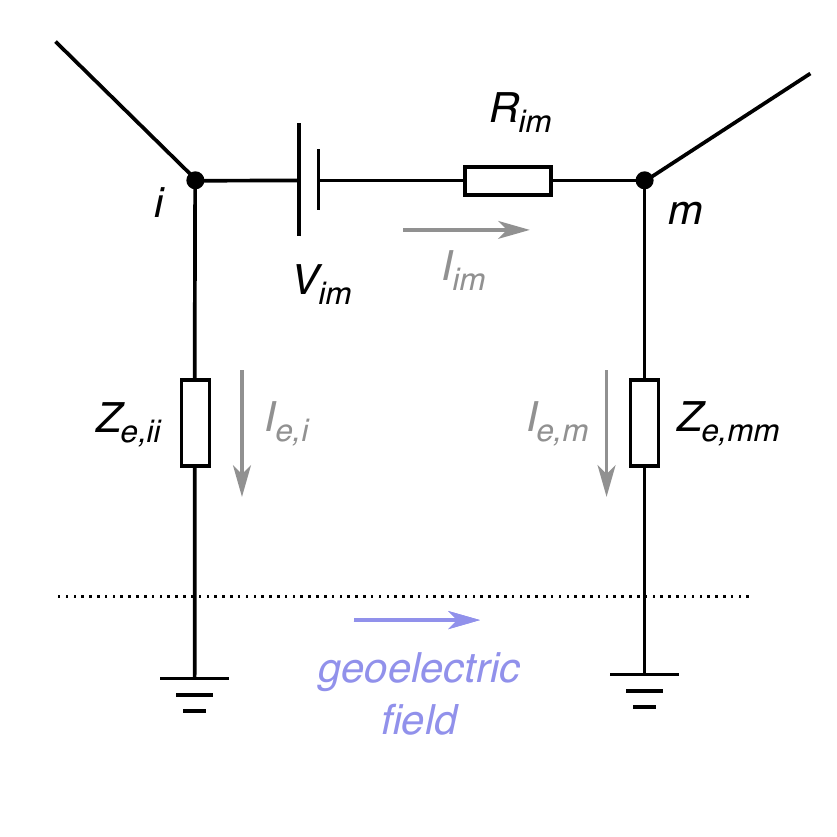}
  \caption{A schematic representation of the equivalent electric circuit between two nodes (that are both connected to the rest of the network) involved in the Lehtinen-Pirjola method.}
  \label{fig:LP-method}
\end{figure}

\subsection{Pirjola-Lehtinen method of GIC modelling}
In the ideal case, the GICs would be measured with sensors intended for this purpose. Unfortunately, these sensors are rare and expensive, so they were installed only at a few points in the power grids with higher risks of GICs-related damages. In the Czech Republic, only one sensor \citep{Ripka2019} was installed recently in substation M\'irovka, unfortunately, the interpretation of the measurements is not straightforward \citep{Hejda2019} and the measurements are not publicly available. 

Several studies from several countries showed that the GICs may successfully be modelled. The modelled values were satisfactorily compared to the direct measurements. The advantage of the GIC modelling is that it is very cheap compared to the development, installation, and maintenance of the GIC sensor and that the modelling may be performed for any situation for which the measurements of the geomagnetic field are available. 

The commonly used method stems from the direct application of Kirchoff's and Ohm's laws when the power-distribution network is virtually replaced by an electric circuit (see Fig.~\ref{fig:LP-method}). In this approach \citep{Lehtinen1985,Pirjola2012}, the network consists of $N$ grounded nodes that are connected by the power lines with a known resistivity. The lines are subjects to the induction by the external electric field. The induced voltages are computed from the known geoelectric field $\boldsymbol{E}$ by 
\begin{equation}
V_{im}=\int_{i}^{m}\boldsymbol{E}\cdot {\rm d}\boldsymbol{s},
\label{eq:Vim}
\end{equation}
where the voltage contributions $\boldsymbol{E}\cdot {\rm d}\boldsymbol{s}$ are integrated along the curve representing the respective line between nodes $i$ and $m$. $V_{im}$ is anti-symmetric to the swapping of $i$ and $m$ due to the opposite orientation of integration paths.

From the known induced voltages, we may compute the ideal-earthing current $J_{e,m}$ in node $m$ from the equation \begin{equation}
J_{e,m}=\sum_{i=1,i\neq m}^{N}\frac{V_{im}}{R_{im}},
\label{eq:J}
\end{equation}
where $V_{im}$ is the geomagnetic voltage calculated from equation (\ref{eq:Vim}) and $N$ is the number of earthed nodes which represent transformer substations of the considered network. $R_{im}$ is the resistance of wires between the nodes $i$ and $m$. Formally, the set of nodes $m=1,\dots,N$ builds a vector of ideal-earthing currents $\boldsymbol{J}_e$, and it represents the currents flowing through the lines under the assumption of the perfect grounding, which will close the circuit. 

Finally, the GICs in the nodes are at once in the vector form given by
\begin{equation}
\boldsymbol{I}_e=(\boldsymbol{1}+\boldsymbol{Y}\cdot\boldsymbol{Z}_e)^{-1}\cdot \boldsymbol{J}_e,
\label{eq:Ie}
\end{equation}
where $\boldsymbol{Y}$ is the admittance matrix and $\boldsymbol{Z}_e$ is the impedance matrix of the whole transmission system. The matrix $\boldsymbol{1}$ indicates an identity matrix $N\times N$. The resulting currents $\boldsymbol{I}_e$ are the sought GICs flowing through the grounding line of the respective node. The convention is such that the positive $I_{e,m}$ indicates the current flowing from the network to the Earth in node $m$, negative $I_{e,m}$ occurs when the current flows from bedrock to the network. We note that usually more than one transformer share the grounding point of the substation. The current flowing through the substation's grounding point is then split between the neutrals of these transformers according to the electrical parameters of the connected transformers. 

To close the set, we give the definitions of the admittance matrix $\boldsymbol{Y}$: 
\begin{equation}
Y_{im}=-\frac{1}{R_{im}}\ {\rm for}\  i \neq m\ {\rm  and}\  Y_{im}=\sum_{k=1,k\neq i}^{N}\frac{1}{R_{ik}}\ {\rm  for}\  i = m,
\end{equation}
where $R_{n,im}$ is the line resistance between two grounded $i$ and $m$ nodes. .\\

The impedance matrix $\boldsymbol{Z}_e$ is defined by the relation
\begin{equation}
Z_{e,im}=R_{i}\ {\rm for}\  i = m\ {\rm  and}\  Z_{e,im}=0\ {\rm  for} \ i \neq m,
\end{equation}
where $R_{i}$ is the grounding resistance of the appropriate node. In the case when the nodes are spatially separated (which is what we assume in our case), the impedance matrix is diagonal.

\section{Results}

Following the above described methodology, we implemented the equations in a code written in {\sc Python} language using packages such as {\sc Numpy}, {\sc Scipy}, {\sc Pandas}, and {\sc Matplotlib}. We took advantage of the structured elements of the language with string-based indexing such as dictionaries or Pandas DataFrames. These structures allowed us to write a universal code, which in principle can be applied to any network topology. Also, the geomagnetic measurements input the code in a form of the external text files (basically in the IAGA2002\footnote{https://www.ngdc.noaa.gov/IAGA/vdat/IAGA2002/iaga2002format.html} format with the header cropped), hence the models may be computed for any geomagnetic field measurements. 

\begin{figure}[h]
  \includegraphics[width=\textwidth]{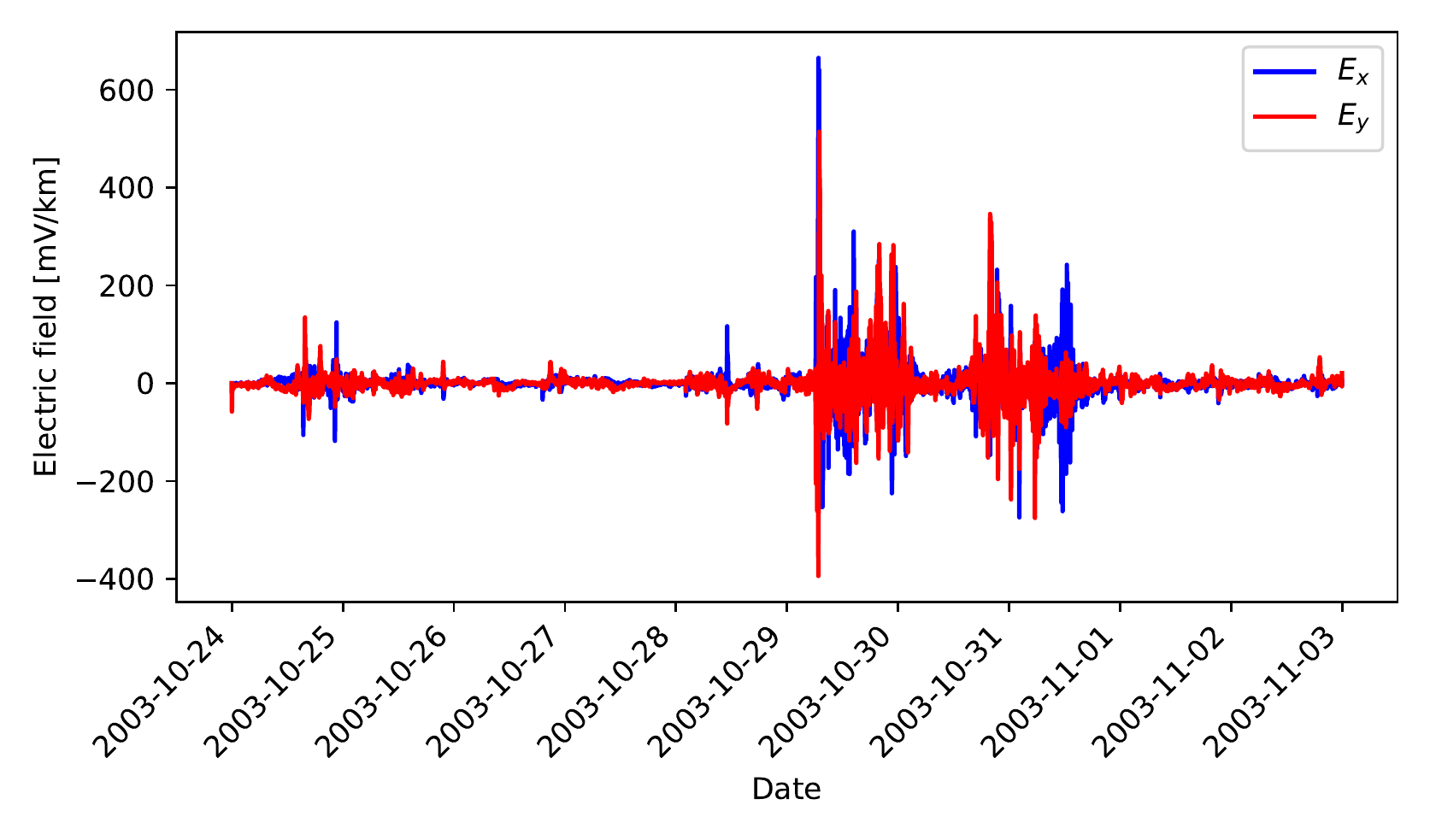}
  \caption{The model for both horizontal components of the geoelectric field during the period of Halloween storms.}
  \label{fig:E}
\end{figure}

\begin{figure}[h]
  \includegraphics[width=\textwidth]{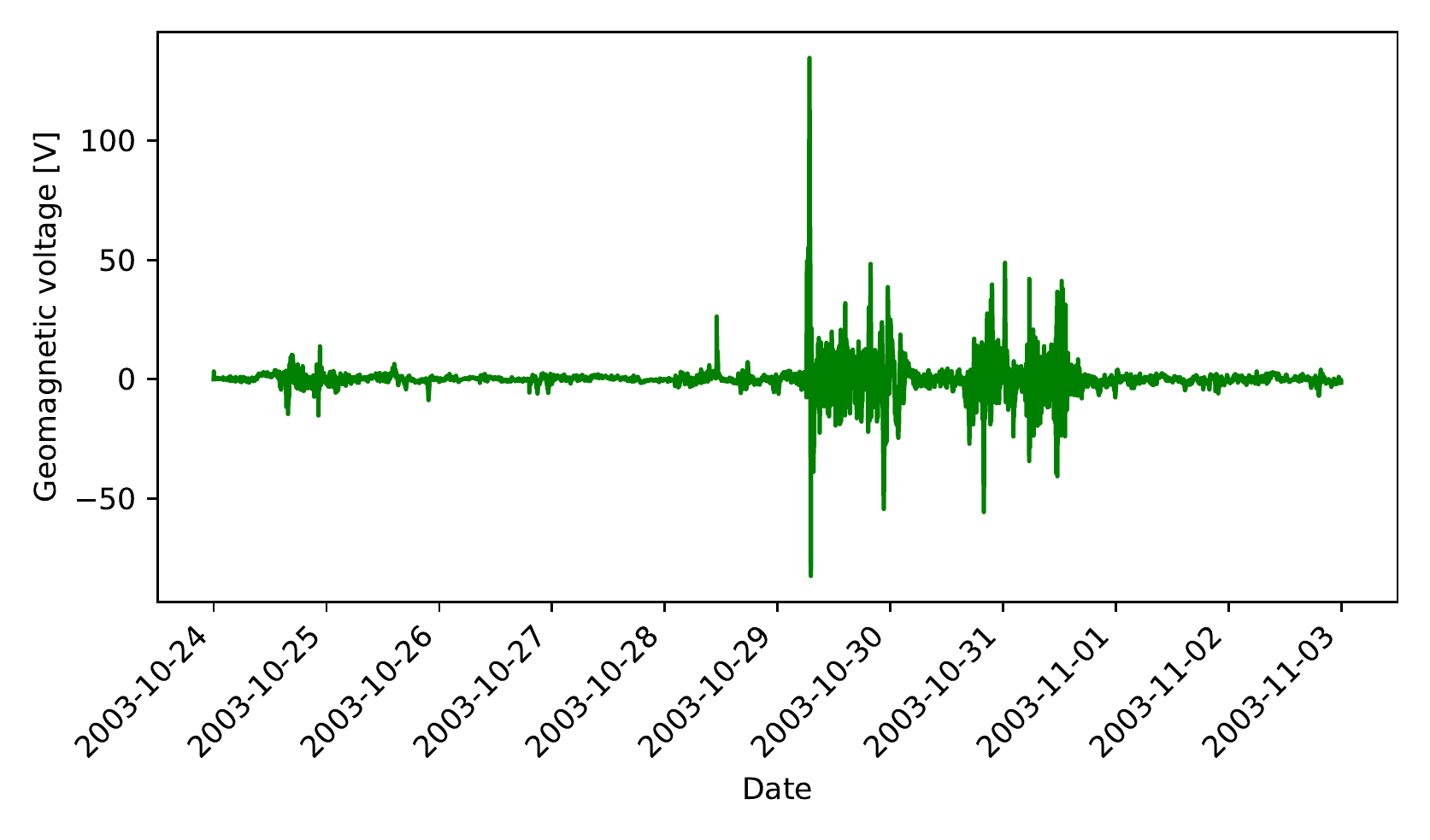}
  \caption{Induced geomagnetic voltage for line V420 during Halloween storms. }
  \label{fig:Vmax}
\end{figure}

\begin{figure}[h]
  \includegraphics[width=\textwidth]{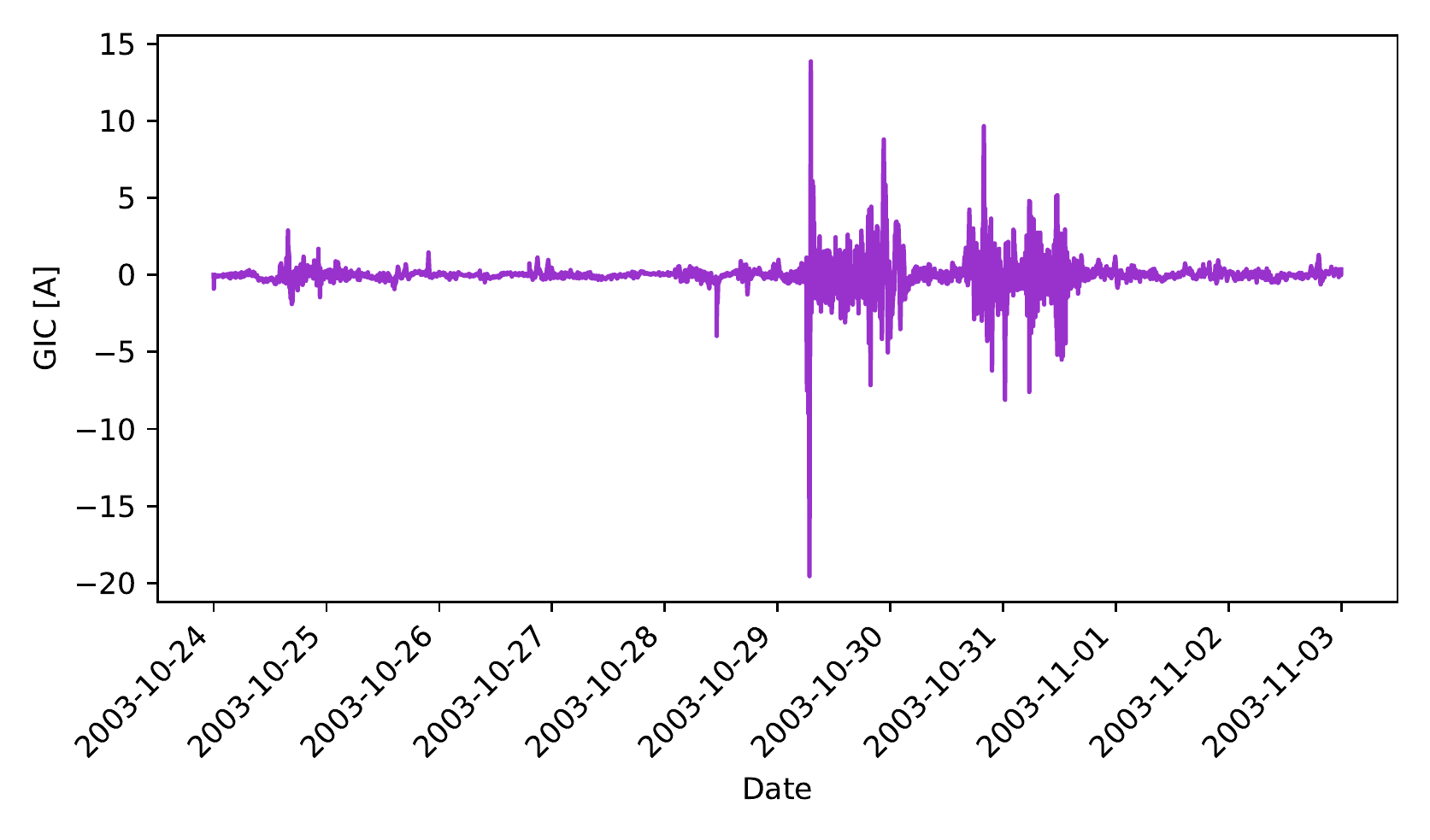}
  \caption{Modelled GIC in substation P\v{r}e\v{s}tice during Halloween storms. The plot indicates the current flowing through the substation's grounding point. Positive values indicate the current flowing from the network to the ground.}
  \label{fig:Iexample}
\end{figure}

\begin{figure}[h]
  \includegraphics[width=\textwidth]{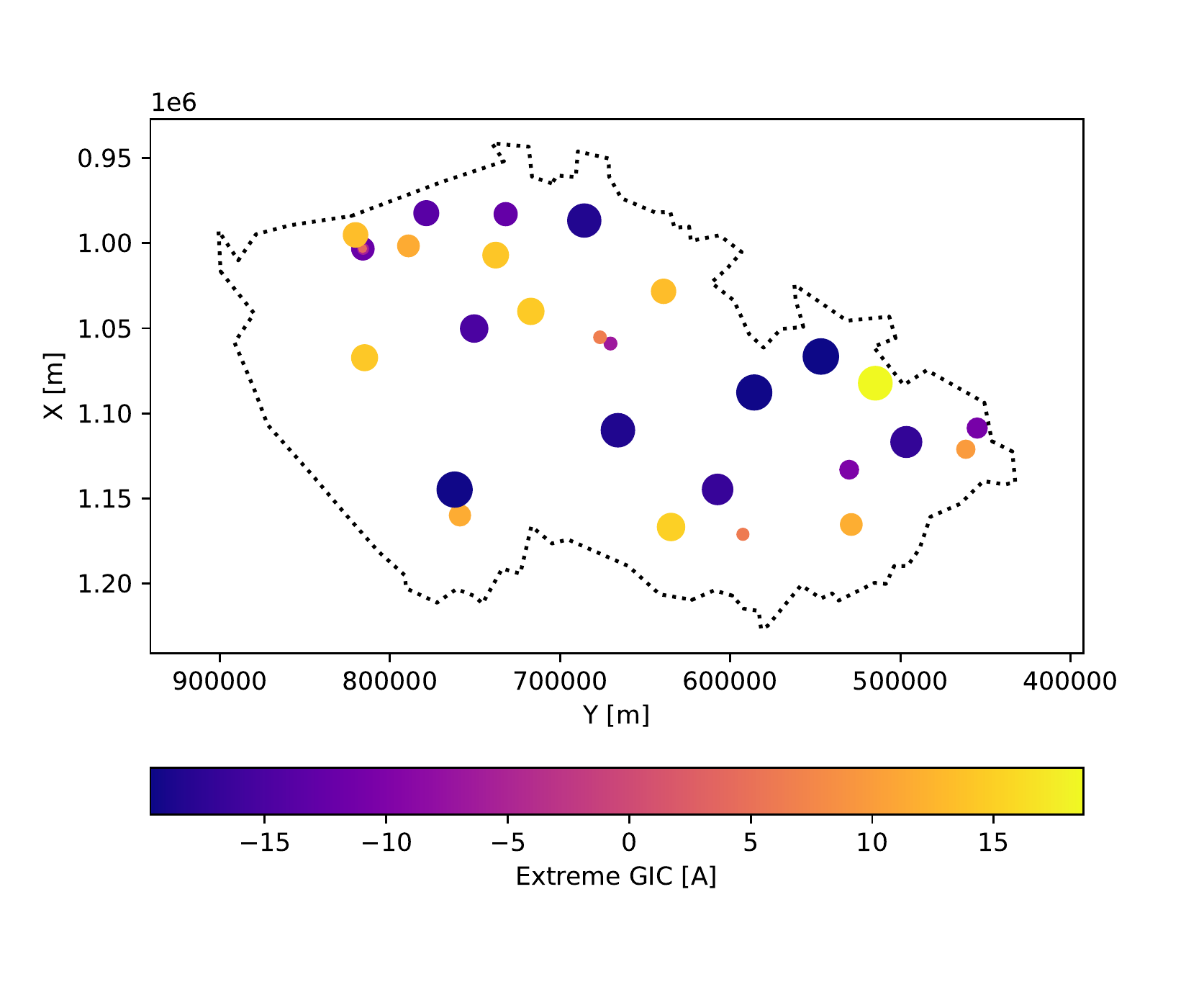}
  \caption{Extreme values of GICs in the investigated substations during the whole period of Halloween storms.}
  \label{fig:Imax}
\end{figure}

\begin{figure}[h]
  \includegraphics[width=\textwidth]{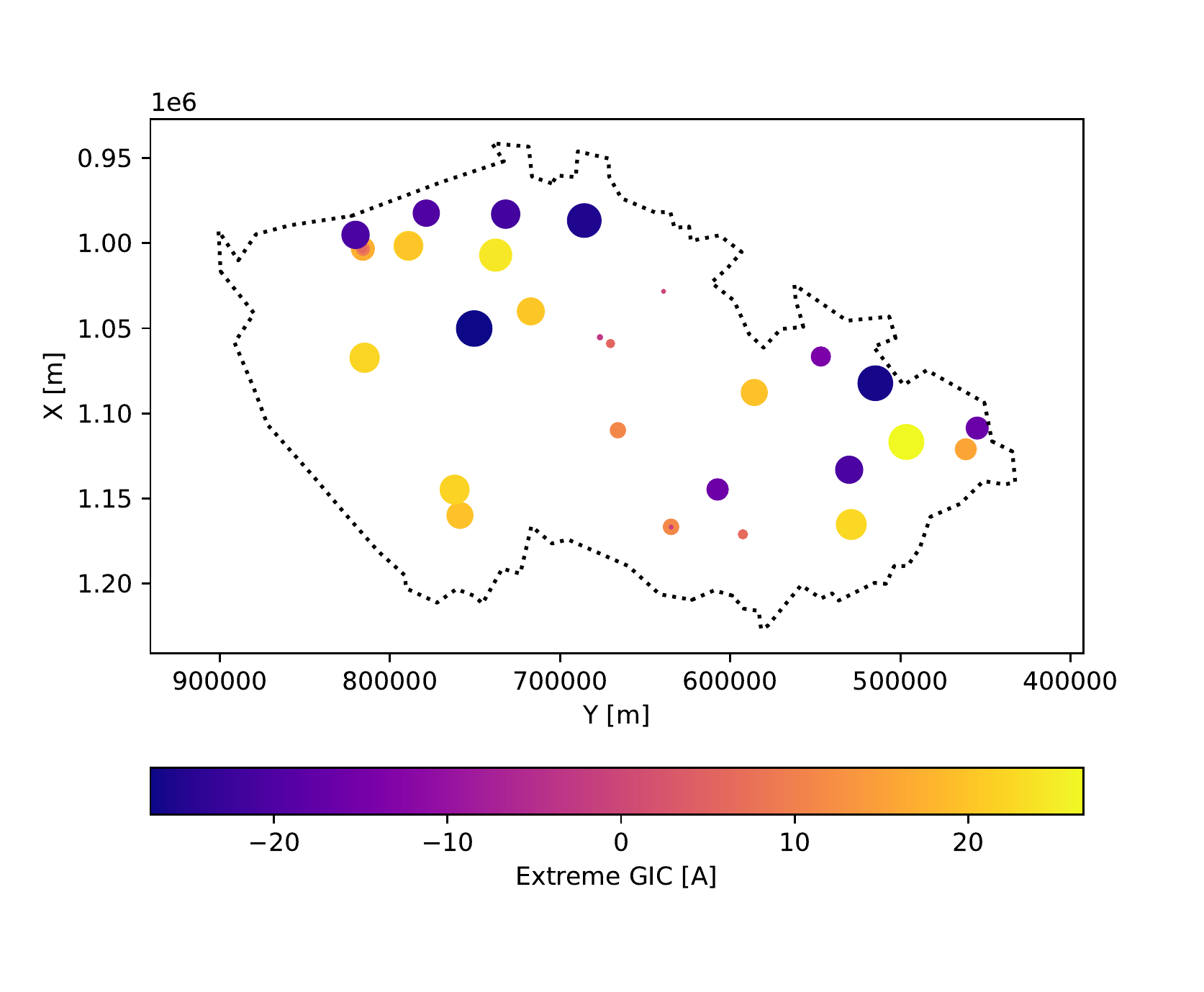}
  \caption{Expected GICs in the studied substations with the 1-V/km northbound electric field.}
  \label{fig:I_1Vkm_N}
\end{figure}

\begin{figure}[h]
  \includegraphics[width=\textwidth]{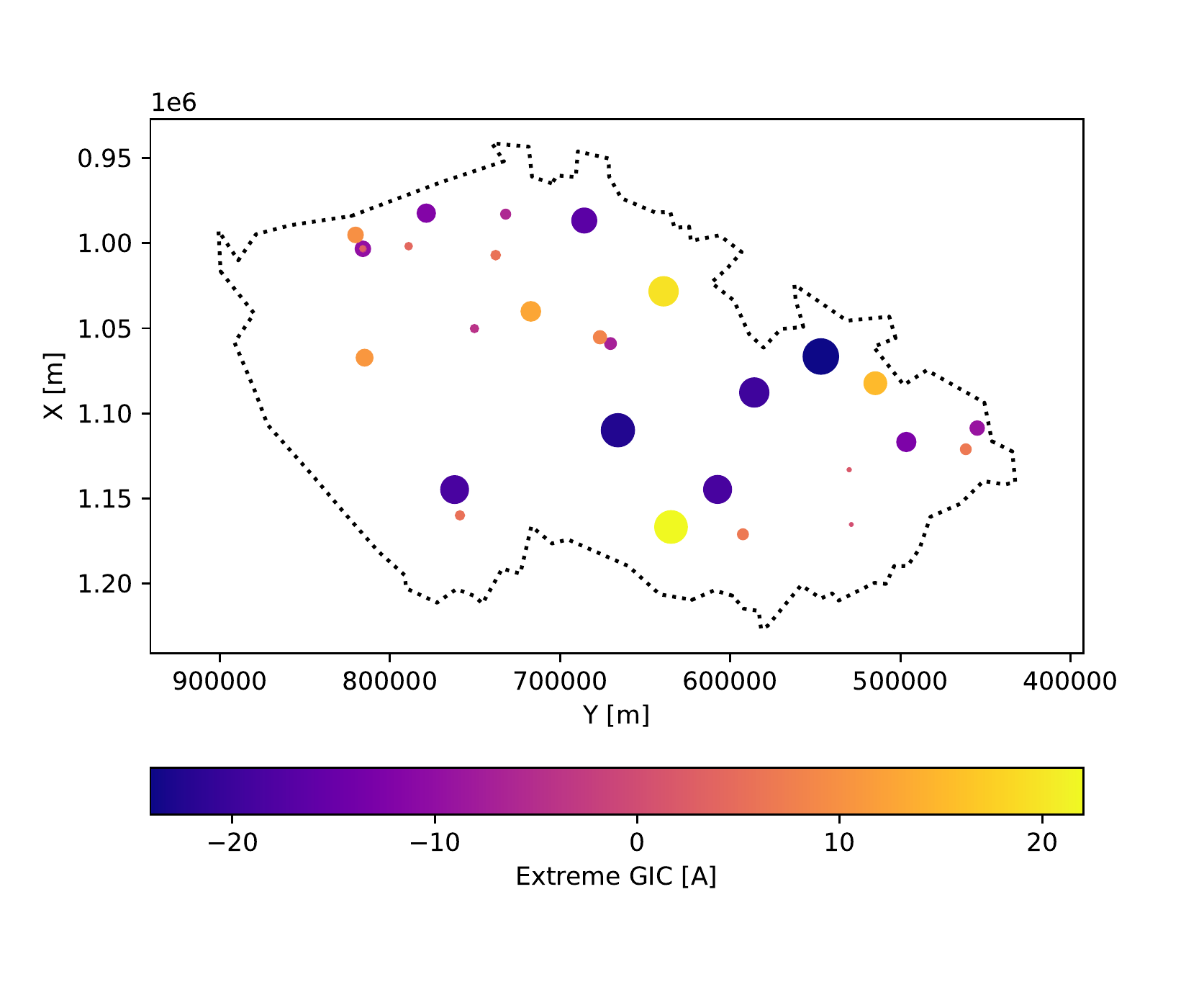}
  \caption{Expected GICs in the studied substations with the 1-V/km eastbound electric field.}
  \label{fig:I_1Vkm_E}
\end{figure}

\begin{figure}[h]
  \includegraphics[width=\textwidth]{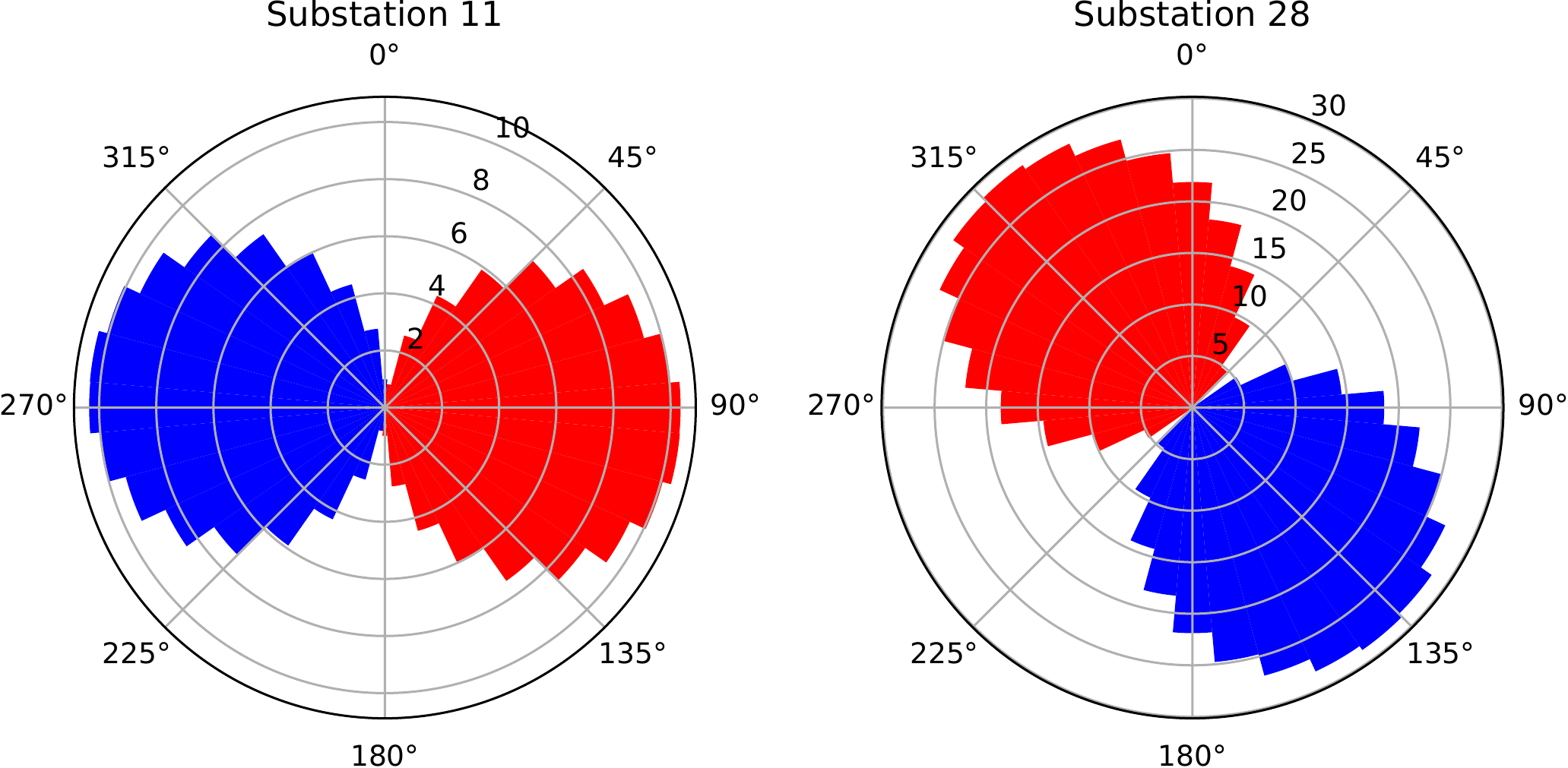}
  \caption{Two examples of the directional analysis for the 1-V/km uniform geoelectric field. The amplitudes of the GIC at the given substation are plotted in the polar plot. The red and blue colours indicate positive and negative GIC values, respectively. The azimuth of 0 indicates the northbound field, value of 90 indicates the eastbound field.}
  \label{fig:I_1Vkm_examples}
\end{figure}

\begin{figure}[h]
 \includegraphics[width=\textwidth]{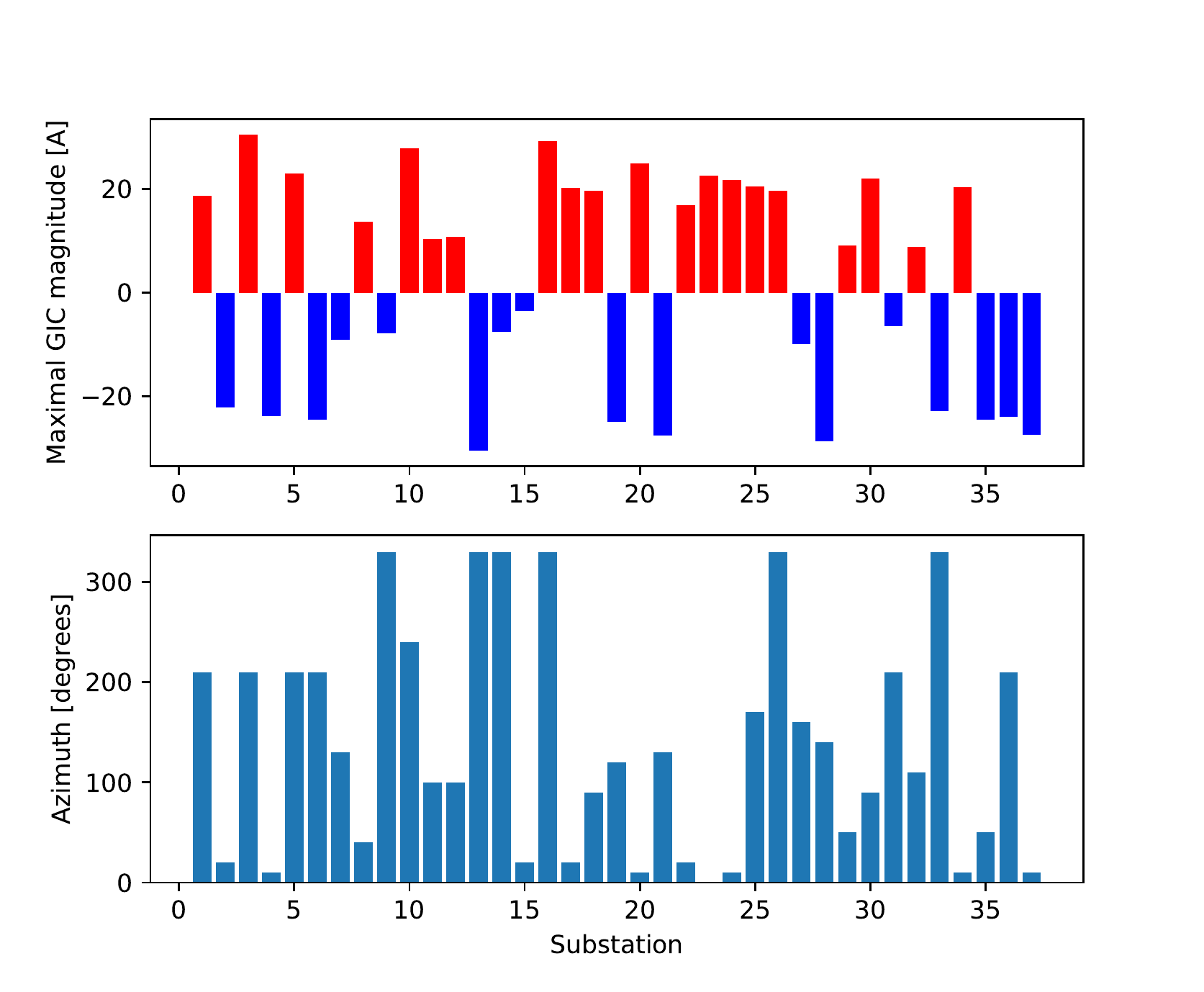}
  \caption{In the upper panel the maximum GICs for the 1-V/km uniform field are given for each considered substation. Red bars indicate positive (from the grid to the soil), blue bar negative values. In the bottom panel the geographical azimuths of the 1-V/km uniform field orientations are given, in which the above given maximum GICs occur.  }
  \label{fig:I_1Vkm_directions}
\end{figure}

\subsection{Halloween storms}

The model of the electric field for the period of the Halloween storms is plotted in Fig.~\ref{fig:E}. The same model was obtained using a similar methodology by \cite{Hejda2005}, which validates our electric field computation code. Using this electric field assuming the plane-wave solution for each 400-kV line, we computed the induced voltages when knowing the resistances of the lines and taking into account the possible multiplicity of the lines. From all 51 lines, the largest voltage was identified for line V420 (see Fig.~\ref{fig:Vmax}), which is about 210~km long and leads from the northwest to nearly the mid-point of the country. Its routing is prevalent in the east-west direction. The induced voltages depend mostly on the length of the line, with maximum values between about 2~V (for 4-km long lines) to several tens of volts for lines with lengths of a few tens kilometres. 

The computed voltages lead to GICs in the order of units or tens amperes. An example GIC flow over the studied period of Halloween storms is given in Fig.~\ref{fig:Iexample} in one substation in Western Bohemia. The largest peaks of about 15~A coincide with the sudden storm commencement, as visible for instance in Fig.~\ref{fig:E}. During the following two days, the GICs changed the sign rapidly with typical amplitudes between one half and one-third of the maximum value. The typical quiet-day GIC amplitudes for this substation are less than 1~A with occasional jumps to about 5~A when the magnetosphere got disturbed temporarily. 

The example shows a typical behaviour of GICs over the studied period for all studied substations in the Czech transmission network. The individual substations differ by the overall scaling of the GIC amplitudes. In Fig.~\ref{fig:Imax} the map of extreme GIC amplitudes over the Halloween storm period for each considered substation is given. This shows that in the studied period, the peak values were at most about 15~A. We note that the extreme values displayed in Fig.~\ref{fig:Imax} did not necessarily occur at the same moment. 

\subsection{Electric field of 1 V/km}

During the Halloween storms, the geomagnetic field was disturbed quite strongly and for a long time. This long-lasting period of geomagnetic storms is considered one of the strongest in recent years. In order to obtain the reasonable estimates of the GIC amplitudes that could possibly be expected in the Czech transmission network, we computed the GIC model for the geoelectric field of the amplitude of 1~V/km, similarly to other studies \citep[e.g.][]{Torta2012,Tozzi2019,Wang2021}. This value is attributed to an extreme geomagnetic storm at middle and low geomagnetic latitudes. Note that during the Halloween storms, the peak magnitudes of the geoelectric field in the Czech Republic were about 650 mV/km, which is about two thirds of the chosen extreme value. 

The resulting induced geovoltages and GICs respectively depend not only on the magnitude of the geoelectric field but also on its direction. For the purpose of this study, we computed two different models. First the model with the 1-V/km geoelectric field pointing towards the north (northbound), the second pointing towards the east (eastbound). 

The expected values in studied substations are given in Fig.~\ref{fig:I_1Vkm_N} for the northbound electric field and in Fig.~\ref{fig:I_1Vkm_E} for the eastbound electric field. Our results show that the expected GIC maxima for the extreme-storm conditions reach values of about 30~A.

The maximum GIC values depend strongly on the uniform field orientation. To investigate this dependence, we additionally computed a set of models with a uniform 1-V/km electric field pointing in various directions (azimuths). Two examples are given in Fig.~\ref{fig:I_1Vkm_examples}. We then indicated the orientation, in which the GIC in each substation separately reached the maximum magnitude. The results of the directional analysis are plotted in Fig.~\ref{fig:I_1Vkm_directions}. There, in the upper panel, the maximum GIC magnitudes for each substation are given, taking the current direction into account. In the bottom panel, we plot the geographical azimuths, in which this maximum GIC amplitude was seen in the models. The inherited azimuth anti-symmetry is clearly visible. The maximum GIC magnitudes reach values of about 30~V. The histogram of the azimuths shows a slight preference for the south-north directions. Together 17 of 37 substations seem to be more susceptible to the geoelectric field oriented not more than 45 degrees from the south-north direction, together 11 substations are more susceptible to the field oriented not more than 45 degrees from the west-east direction. 

\section{Discussion}

The results of the GIC modelling in the Czech transmission network show that the GIC amplitudes in substations during strong geomagnetic storms do not reach the values that could lead to catastrophic failure of the network devices. On the other hand, for instance, \cite{Koen2003} showed that even GIC levels as low as 1--10~A may lead to magnetic core saturation of the exposed transformer and produce extremely large and highly distorted AC to be drawn from the power grid. This amplified AC from saturation effects can pose risks to the power networks directly due to increased reactive power demands that can cause voltage regulation problems. 

Our recent statistical studies \citep{Vybostokova2019,Svanda2020,VybostokovaPREP} together with statistical studies from other countries \citep[e.g.][]{Schrijver2013,Zois2013,Schrijver2014,Gil2019,Gil2021} demonstrate that the increased geomagnetic activity statistically leads to a short-term increase of anomaly rates registered in the maintenance logs. The connection in the statistical studies is always somewhat loose. Repeated exposure of the power-grid elements to small GICs may disturb their function, which, in general, will lead not to catastrophic failure, but to something that will appear as ``ageing'' to the maintainer. 

There are several simplifications we used in our study that might influence the accuracy of the derived GIC values. We considered a simple plane-wave model of the geoelectric field with only a single conductivity value representative for the territory of the whole country. It certainly would be more proper to use a more sophisticated model for instance by using the 1-D conductivity EURHOM model \citep{Adam2012} used e.g by the EURISGIC project \citep{Viljanen2012}. On the other hand, \cite{Viljanen2004} pointed out that the plane-wave models (of the geoelectric field) are fairly accurate for GIC purposes. They arrived at this conclusion when testing the performance of several methods deriving the models of the geoelectric field. Furthermore, we need to stress out that we used the same model as \cite{Hejda2005}, where the modelled voltages (using the plane-wave model) were directly compared to the measured pipe-to-soil voltages with great success (see their Fig.~6). Thanks to this work the simple plane-wave model may be considered as validated to real measurements and thus accurate enough for our purpose. 

One needs to keep in mind that in our study there are more potential caveats than the model of the geoelectric field. The model of the geomagnetic field is also considered uniform, because we use the measurements from only a single station at Budkov. The territory of the Czech Republic is rather small; however, in principle, single-point measurements cannot cover possible spatial variations. This is further supported by the findings of \cite{Viljanen2004}, who pointed out that the proper representation of the geomagnetic field is more crucial for the determination of the GICs than possible spatial variations of the geoelectric field. Also, some of the technical parameters of the model of the power grid are certainly idealised. It is unlikely that the 100-km long power line will have ideally homogeneous and stable parameters (such as the resistivity) in real conditions. The possible deviations are unknown. The inaccuracies in the power-grid model may potentially have a very large effect on the exact values of the computed GICs.   

In the Pirjola-Lehtinen model, the GIC (the current flowing between the Earth and the node) amplitudes depend on the values of the grounding resistances, see equation (\ref{eq:Ie}). Unfortunately, in our case, the grounding resistances were known only for a few substations. For the remaining substations, we chose the average of the known values, that is 0.1~$\Omega$. This is not an unreasonable value. For instance, \cite{Torta2014} faced a similar problem and chose the value of 0.15~$\Omega$. The values in the order of 0.1~$\Omega$ are seen also in other papers dealing with the GIC modelling.   

Since the solution depends on the chosen value, we tested its sensitivity on the value chosen. Apart from the favoured solution, we thus computed two more solutions for the unknown grounding resistances having the maximum value of the known ones (that it 0.272~$\Omega$) and the minimum value from the known ones (that is 0.031~$\Omega$). We note that for the substations with the known grounding resistances, we always kept the known values. 

The sensitivity was tested by comparing the GIC values for the two supplementary solutions with the favourable one. We found that for the substations, for which the grounding resistances were known, the resulting GICs did not change much with the change of the resistivities of the remaining nodes. Over the period of Halloween storms, the differences between the solutions in the substations with the fixed grounding resistances are less than 0.1~A. This indicates that the GIC amplitude computed for the substation with a fixed grounding resistance is only weakly influenced by grounding resistances of other nodes in the power grid. 

On the other hand, for substations with unknown grounding resistance, the differences are more significant. The GIC amplitudes may consequently be different even by 50 per cent or more in some peaks and, to the first approximation, they are inversely proportional to the grounding-resistance value. The histograms of the differences between the different solutions are considerably wider than in the case of substations with known resistances. Still, the histograms are strongly peaked around the bin with the value of the difference less than 0.1~A, hence the solutions using different earthing resistances are still somewhat comparable. 

From the sensitivity analysis it seems that knowing all grounding resistances of the considered substations will make the results more precise, however, the estimate by some generic value does not invalidate the results. Our findings are similar to those discussed by \cite{Viljanen2013}.

Our estimated peak amplitudes for the period of Halloween storms are comparable to a similar analysis performed for neighbouring Austria. \cite{Bailey2017} estimated the GIC amplitudes to about 13~A. On the other hand, the Europe-wide model by \cite{Wik2009} predicted peak values of about 50~A, which is about three times larger than values obtained by us. We assume that the incomplete model of the transmission grid, when only the spine was considered, plays a considerable role.  

Lastly, we have to admit a small inconsistency in our modelling. We modelled the expected GIC amplitudes during the period of Halloween storms in 2003, however, we used the network topology as of end 2019. Between 2003 and 2019, several grid modifications were made, including the splitting of some lines into two. We do not consider this inconsistency very important. Our study is devoted to show the expected GIC amplitudes in the today's grid, and we only used the period of Halloween storms as an example of what could be expected from our Sun. 

\section{Conclusions}

Investigation and assessment of the natural hazards belong to the disciplines, where exact science may contribute significantly. The hazards related to solar activity have been in serious focus since about 20 years ago. Their investigation becomes routine even in countries, where these effects were considered non-existent. The assessment of possible GIC amplitudes in the mid- and low-latitude country is thus very timely. 

According to our modelling, the peak GIC amplitudes in a single node that can be expected in strong geomagnetic storms are 30~A at most. This is a much too low value to pose an immediate threat to transformers or other grid devices or elements. On the other hand, the repeated exposure to GICs of this magnitude may fasten the ageing of the devices and thus shorten their lifetime. That would explain the statistically significant increase of the anomaly rates after the exposure to solar activity seen in many studies. 


\begin{backmatter}

\section*{Abbreviations}
ACE = Advanced Composition Explorer\\
CIR = Co-rotating Interaction Region\\
CME = Coronal Mass Ejection\\
EURISGIC = European Risk from Geomagnetically Induced Currents\\
EURHOM = European Rho Model\\
GIC = Geomagnetically Induced Current\\
GPS = Global Positioning System\\

\section*{Availability of data and materials}

The geomagnetic measurements necessary to compute the geoelectric field are available from within the International Real-time Magnetic Observatory Network (INTERMAGNET), which can be downloaded from \mbox{https://intermagnet.github.io/}. 

The technical data of the Czech transmission network that support the findings of this study are available from \v{C}EPS, a.s. (\mbox{http://www.ceps.cz/en/homepage}), but restrictions apply to the availability of these data, which were used under license for the current study, and so are not publicly available. Data are however available from the authors upon reasonable request and with permission of \v{C}EPS, a.s.

\section*{Competing interests}

The authors declare no competing interests.

\section*{Funding}
ASU CAS is funded by the institute research project ASU:67985815. 

\section*{Authors' contributions}

This paper presents the results of two master theses. M\v{S} designed the research, supervised both AS and TV and wrote the initial draft of the paper. AS communicated with the data providers, wrote the GIC modelling code, and processed and analysed the data. TV wrote the code to compute the model for the geoelectric field from geomagnetic field measurements. All authors contributed to the final manuscript.

\section*{Acknowledgments}
The results presented in this paper rely on the data collected at Budkov. We thank the Institute of Geophysics of the CAS, for supporting its operation and INTERMAGNET for promoting high standards of magnetic observatory practise (www.intermagnet.org). We thank two anonymous reviewers for their reports which helped to improve the quality of the paper. 

\bibliographystyle{bmc-mathphys} 
\bibliography{biblio}      
\nocite{label}

\end{backmatter}
\end{document}